\documentclass[aps,preprint,showpacs,floats,epsf,epsfig,nofootinbib,12pt]{revtex4}
\usepackage{graphicx}
\usepackage{epsfig}
\usepackage[dvips]{color}

\def\beq{\begin{eqnarray}}
\def\eeq{\end{eqnarray}}
\def\non{\nonumber}

\def\la{\langle}
\def\ra{\rangle}
\def\Mb{M_{\mathcal{B}_1}}
\def\Mc{M_{\mathcal{B}_2}}

\begin{document}

\title{The weak decays of $\Xi^{(')}_{c}\to\Xi$ in the light-front quark model}

\vspace{1cm}

\author{ Hong-Wei Ke$^{1}$   \footnote{khw020056@tju.edu.cn}, Qing-Qing Kang$^{1}$, Xiao-Hai Liu$^{1}$ \footnote{xiaohai.liu@tju.edu.cn} and
        Xue-Qian Li$^2$\footnote{lixq@nankai.edu.cn},
   }

\affiliation{  $^{1}$ School of Science, Tianjin University, Tianjin 300072, China
\\
  $^{2}$ School of Physics, Nankai University, Tianjin 300071, China }

\vspace{12cm}

\begin{abstract}

Without contamination from the final state interactions, the
calculation of the branching ratios of semileptonic decays
$\Xi^{(')}_{c}\to\Xi+e^+\nu_e$ may provide us more information about
the inner structure of charmed baryons. Moreover, by studying those
processes, one can better determine the form factors of
$\Xi_c\to\Xi$ which can be further applied to relevant estimates. In
this work, we use the light-front quark model to carry out the
computations where the three-body vertex functions for $\Xi_c$ and
$\Xi$ are employed. To fit the new data of the Belle II, we
re-adjust the model parameters and obtain  $\beta_{s[sq]}=1.07$ GeV
which is 2.9 times larger than $\beta_{s\bar s}=0.366$ GeV. This
value may imply that the $ss$ pair in $\Xi$ constitutes a more
compact subsystem. Furthermore, we also investigate the non-leptonic
decays of $\Xi^{(')}_c\to \Xi$ which will be experimentally measured
soon, so our model would be tested by consistency with the new data.

 \pacs{13.30.-a,12.39.Ki, 14.20.Lq}
\end{abstract}

\maketitle

\section{Introduction}

Recently the Belle Collaboration measured the branching fraction of
the semi-leptonic decay of the charmed baryon $\Xi_c$ as
$\mathcal{B}({\Xi^0}_c\to \Xi^-
e^+\nu_e)=(1.72\pm0.10\pm0.12\pm0.50)\%$\cite{Li:2021uhk} where the
first and second uncertainties are statistical and systematic
whereas the third one arises from the uncertainty of
$\mathcal{B}({\Xi^0}_c\to \Xi^- \pi^+)=(1.80\pm0.52) \%$ measured
also by the Belle collaboration\cite{Li:2018qak}. In 1993 the ARGUS
Collaboration\cite{Albrecht:1992jt} first observed the decay
${\Xi^0}_c\to \Xi^- e^+\nu_e$  and in 1995 the CLEO
Collaboration\cite{Alexander:1994hp} also found ${\Xi^0}_c\to \Xi^-
e^+\nu_e$ events and the ratio $\mathcal{B}({\Xi^0}_c\to \Xi^-
e^+\nu_e)/\mathcal{B}({\Xi^0}_c\to \Xi^- \pi^+)$ has been measured
to be $0.96\pm0.43\pm0.18$ by the ARGUS and
$3.1\pm1.0^{+0.3}_{-0.5}$ by the CLEO, respectively. With the data
$\mathcal{B}({\Xi^0}_c\to \Xi^- \pi^+)=(1.80\pm0.52)\%$  the
previous average
 $\mathcal{B}({\Xi^0}_c\to \Xi^- e^+\nu_e)$ was $(2.34\pm 1.59)\%$\cite{Zyla:2020zbs}.
 Apparently the measurement of the Belle Collaboration on $\mathcal{B}({\Xi^0}_c\to \Xi^- e^+\nu_e)$
is more precise at present. Recently a lattice calculation on
$\mathcal{B}({\Xi^0}_c\to \Xi^- e^+\nu_e)$ has been finished in
Ref.\cite{Zhang:2021oja} and the result is $(2.29\pm0.29\pm0.31)\%$.
Some theoretical predictions based on different phenomenological
models are given as $(3.4\pm0.7)\%$\cite{Zhao:2021sje},
$(3.49\pm0.95)\%$\cite{Geng:2020gjh},$(7.26\pm2.54)\%$\cite{Azizi:2011mw},
$1.35\%$\cite{Zhao:2018zcb}, $(4.87\pm1.74)\%$\cite{Geng:2018plk},
$(2.4\pm0.3)\%$\cite{Geng:2019bfz}, and
$2.38\%$\cite{Faustov:2019ddj}. Definitely the precision of
theoretical evaluations should be further improved.

 In this work we will employ the light-front quark model to study
the weak decays of $\Xi^{(')}_{c}\to \Xi$.  The light-front quark
model (LFQM) is a relativistic quark model which has been applied to
study transitions among mesons and the relevant theoretical
predictions with this model agree with the data within reasonable
error tolerance
\cite{Jaus,Ji:1992yf,Cheng:1996if,Cheng:2003sm,Hwang:2006cua,Lu:2007sg,
Li:2010bb,Ke:2009ed,Ke:2010,Wei:2009nc,Choi:2007se,
Ke:2009mn,Ke:2011fj,Ke:2011mu,Ke:2011jf}. Later the model was
extended to study the decays of pentaquark within the
diquark-diquark-antiquark picture\cite{pentaquark1,pentaquark2} and
the weak decay of baryons in the quark-diquark
picture\cite{Ke:2007tg,Wei:2009np,Ke:2012wa,Ke:2017eqo,Wang:2017mqp,Chua:2018lfa,Yu:2017zst}.
With this model we employed the three-body vertex function to
explore the decays of $\Lambda_b$ and $\Sigma_b$\cite{Ke:2019smy}
where three individual quarks are concerned, the related picture is
somewhat different from the structure where one-diquark and
one-quark stand as basic constituents. Geng et.
al.\cite{Geng:2020gjh} also studied the decay of baryons under the
three-quark picture in the light front quark model where they took
different approaches from ours to deal with the flavor-spin wave
function of baryons. It is noted that in the three-body structure,
even though two quarks might be loosely bound as a subsystem with
definite spin and color, unlike the diquark which is a relatively
stable subject, the subsystem may break and its components would
undergo a re-combination with other quarks to constitute different
sub-systems, especially after a hadronic transition.

In the three-body vertex function of baryon two quarks join together
into a subsystem which has a definite spin and then the subsystem
couples with the rest quark to form a baryon with the defined spin.
In order to evaluate the transition between two baryons we first
need to know the inner spin structures of the concerned baryons. For
single charmed-baryons $\Xi_c$ and $\Xi'_c$, the two light quarks
$sq$ where $q$ denotes $u$ or $d$ quark can be regarded as a
subsystem with definite spin (in analog to a diquark
)\cite{Ebert:2006rp,Korner:1992wi}. By contrast, there are three
light quarks in $\Xi$ where two $s$ quarks possess definite spin 1
due to the antisymmetry of total wave function, so that can be
regarded as a subsystem. In the transition of $\Xi^{(')}_{c}\to\Xi$
the $c$ quark in the initial state would transit into an $s$ quark
by emitting a gauge boson and the original $sq$ pair can be regarded
as a spectator which does not undergo any changes during the
hadronic transition. However in the final state the newly emerged
quark $s$  would couple to the $s$ quark which is the decay product
of the initial $c$ quark to form a physical $ss$ subsystem with
definite spin, thus the $sq$ from initial state is no longer a
physical subsystem. Namely the picture is that the old subsystem
$sq$ is broken and a new subsystem $ss$ emerges during the hadronic
transition. Definitely, we would account the changes of constituents
in the subsystems as caused by a non-perturbative QCD effect.
Anyhow, the $ss$ subsystem isn't a spectator because one of the $s$
quarks originates from weak decay of the initial $c$ quark.
Therefore the simple quark-diquark picture does not apply to these
processes. In order to use the spectator proximation, one needs to
rearrange the ($ss$)-($q$) structure into the $s$-($sq$) structure
by a Racah transformation.

 This paper is
organized as follows: after the introduction, in section II we
deduce the transition amplitude for $\Xi^{(')}_{c}\rightarrow \Xi$
in the light-front quark model and provide the expressions of the
form factors, then we present our numerical results for
$\Xi^{(')}_{c}\rightarrow \Xi$ in section III. Section IV is
devoted to the conclusion and discussions.

\section{$\Xi_{c}\rightarrow \Xi$ and $\Xi'_{c}\rightarrow \Xi$  in the light-front quark model}

\subsection{The vertex functions of $\Xi_{c}$, $\Xi'_{c}$ and $\Xi$}
Enlightened by Ref.\cite{Tawfiq:1998nk} in our previous
work\cite{Ke:2019smy} we constructed the vertex functions of baryons
under the three-quark picture where two quarks constitute a
subsystem with definite spin, and then the subsystem couples with
the rest quark to form a baryon. We have employed the vertex
functions to study the decays of $\Lambda_b$ and
$\Sigma_b$\cite{Ke:2019smy}. Later we employed the three-quark
vertex functions to study the transition $\Xi_{cc}\to
\Xi_c$\cite{Ke:2019lcf}. The success encourages us to apply this
picture to further study relevant processes.

In analog to Refs.\cite{Ke:2019smy}  the vertex functions of
$\Xi_{c}$, $\Xi'_{c}$ and $\Xi$  with total spin $S=1/2$ and
momentum $P$ are
\begin{eqnarray}\label{eq:lfbaryon}
&& |\Xi_{c}^{(')}(P,S,S_z)\rangle=\int\{d^3\tilde p_1\}\{d^3\tilde
p_2\}\{d^3\tilde p_3\} \,
  2(2\pi)^3\delta^3(\tilde{P}-\tilde{p_1}-\tilde{p_2}-\tilde{p_3}) \nonumber\\
 &&\times\sum_{\lambda_1,\lambda_2,\lambda_3}\Psi^{SS_z}_{\Xi_{c}^{(')}}(\tilde{p}_i,\lambda_i)
  \mathcal{C}^{\alpha\beta\gamma}\mathcal{F}_{csq}\left|\right.
 c_{\alpha}(p_1,\lambda_1)s_{\beta}(p_2,\lambda_2)q_{\gamma}(p_3,\lambda_3)\ra,\\
  && |\Xi(P,S,S_z)\rangle=\int\{d^3\tilde p_1\}\{d^3\tilde
p_2\}\{d^3\tilde p_3\} \,
  2(2\pi)^3\delta^3(\tilde{P}-\tilde{p_1}-\tilde{p_2}-\tilde{p_3}) \nonumber\\
 &&\times\sum_{\lambda_1,\lambda_2,\lambda_3}\Psi^{SS_z}_{\Xi}(\tilde{p}_i,\lambda_i)
  \mathcal{C}^{\alpha\beta\gamma}\mathcal{F}_{ssu}\left|\right.
  s_{\alpha}(p_1,\lambda_1)s_{\beta}(p_2,\lambda_2)q_{\gamma}(p_3,\lambda_3)\ra.
\end{eqnarray}


Let us repeat some details about the three-body picture because it
plays the key role in our calculations. In order to obtain the
expression of $\Psi^{SS_z}_{\Xi_{c}^{(')}}$ and $\Psi^{SS_z}_{\Xi}$
one needs to know their  inner spin-flavor structure. In
Ref.\cite{Ebert:2006rp} the $sq$ in $\Xi_{c}$ is considered as a
scalar subsystem whereas in $\Xi'_{c}$ it is a vector. In the decay
process $\Xi^{(')}_{c}\to\Xi$ the $c$ quark transits into an $s$
quark via weak interaction and the original $sq$ subsystem can be
approximately regarded as a spectator because it does not undergo
any change during the hadronic transition. However the two strange
quarks in $\Xi$ compose a physical subsystem whose spin is 1. To
apply the spectator approximation for the transition, we rearrange
the quark structure of $(ss)$-$q$ into a sum of $\sum_i s$-$(sq)_i$
where $i$ runs over all possible spin projections by a Racah
transformation. With the rearrangement of quark spin-flavor the
physical structure $(ss)$-$q$ in $\Xi$ is rewritten into a sum over
the effective structures of $s$-$(sq)$. The detailed transformations
are\cite{Wang:2017mqp}
\begin{eqnarray}\label{RK1}
&&{[s^1s^2]}_{1}[q]=-\frac{\sqrt{3}}{2}[s^1][s^2q]_{0}+
\frac{1}{2}[s^1][s^2q]_{1},
\end{eqnarray}
and then
\begin{eqnarray}\label{RK2}
\Psi^{SS_z}_{\Xi}(\tilde{p}_i,\lambda_i)=-\frac{\sqrt{3}}{2}\Psi^{SS_z}_0(\tilde{p}_i,\lambda_i)
+\frac{1}{2}\Psi^{SS_z}_1(\tilde{p}_i,\lambda_i),
\end{eqnarray}
with\cite{Tawfiq:1998nk,Ke:2019smy}

\begin{eqnarray}
\Psi^{SS_z}_0(\tilde{p}_i,\lambda_i)=&&A_0 \bar
U(p_3,\lambda_3)[(\bar
P\!\!\!\!\slash+M_0)\gamma_5]V(p_2,\lambda_2)\bar U(p_1,\lambda_1)
U(\bar P,S) \varphi(x_i,k_{i\perp}),\\
A_0=&&\frac{1}
{4\sqrt{P^+M_0^3(m_1+e_1)(m_2+e_2)(m_3+e_3)}},\nonumber
\end{eqnarray}
\begin{eqnarray}
\Psi^{SS_z}_1(\tilde{p}_i,\lambda_i)=&&A_1 \bar
U(p_3,\lambda_3)[(\bar
P\!\!\!\!\slash+M_0)\gamma_{\perp\alpha}]V(p_2,\lambda_2)\bar
U(p_1,\lambda_1) \gamma_{\perp\alpha}\gamma_{5}U(\bar
P,S)\varphi(x_i,k_{i\perp}),\\
A_1&&=\frac{1}{4\sqrt{3P^+M_0^3(m_1+e_1)(m_2+e_2)(m_3+e_3)}},\nonumber
\end{eqnarray}
where $p_1$ is the the momentum of the newly emerged $s$ quark in
the transition, $\,p_2\,,p_3$ are the momenta of the spectator
quarks $s$ and $q$, $U$ and $V$ are spinors and
$\lambda_1,\lambda_2, \lambda_3$ are the helicities of the
constituents. Since the spin of $sq$ subsystem is 0 (1), the
expressions of $\Psi^{SS_z}_{\Xi_{c}}$ (
$\Psi^{SS_z}_{\Xi_{c}^{'}}$) is the same as $\Psi^{SS_z}_0$
($\Psi^{SS_z}_1$) except that $p_1$ is the momentum of the $c$
quark.

The spatial wave function is
 \beq\label{A122}
\varphi(x_i,k_{i\perp})=\frac{e_1e_2e_3}{x_1x_2x_3M_0}\varphi(\overrightarrow{k}_1,\beta_1)
\varphi(\frac{\overrightarrow{k}_2-\overrightarrow{k}_3}{2},\beta_{23}),
 \eeq
with
$\varphi(\overrightarrow{k},\beta)=4(\frac{\pi}{\beta^2})^{3/4}{\rm
exp}(\frac{-k_z^2-k^2_\perp}{2\beta^2})$.

\subsection{Calculating the  form factors of $\Xi^{(')}_{c}\to\Xi_c$ in LFQM}

\begin{figure}
\begin{center}
\scalebox{0.6}{\includegraphics{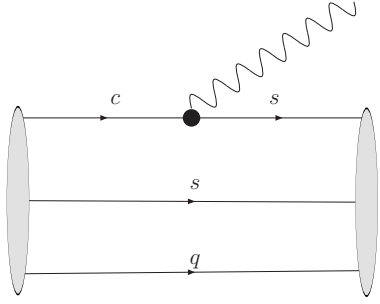}}
\end{center}
\caption{The Feynman diagram for $\Xi^{(')}_{c}\to\Xi$ transition,
where $\bullet$ and $q$ denote $V-A$ current vertex and $u$ or
 $d$ quark respectively.}\label{t1}
\end{figure}

First we consider the weak decay $\Xi_{c}\to\Xi$. The leading order
Feynman diagram is depicted in Fig. \ref{t1}. Following the
procedures given in
Ref.\cite{pentaquark1,pentaquark2,Ke:2007tg,Ke:2012wa} the
transition matrix element can be evaluated with the vertex
functions of $\mid \Xi_{c}(P,S,S_z) \ra$ and $\mid
\Xi^(P',S',S'_z) \ra$. The $sq$ subsystem stands as a spectator,
i.e. its spin-flavor configuration does not change during the
transition, so only the first term in Eq.(\ref{RK2}) can contribute to the transition. The hadronic matrix element can be:

\begin{eqnarray}\label{s00}
&& \la \Xi(P',S_z') \mid \bar{s}
\gamma^{\mu} (1-\gamma_{5}) c \mid \Xi_{c}(P,S_z) \ra  \nonumber \\
 &=&-\frac{\sqrt{3}}{2} \int\frac{\{d^3 \tilde p_2\}\{d^3 \tilde p_3\}\phi_{\Xi}^*(x',k'_{\perp})
  \phi_{\Xi_{c}}(x,k_{\perp})Tr[(\bar{P'}\!\!\!\!\!\slash'-M_0')\gamma_{5}(p_2\!\!\!\!\!\slash+m_2)
 (\bar{P}\!\!\!\!\!\slash'+M_0)\gamma_{5}(p_3\!\!\!\!\!\slash-m_3)]}{16\sqrt{p^+_1p'^+_1\bar{P}^+\bar{P'}^+M_0^3(m_1+e_1)
 (m_2+e_2)(m_3+e_3)(m_1'+e_1')
 (m_2'+e_2')(m_3'+e_3')}}\nonumber \\
  &&\times  \bar{u}(\bar{P'},S'_z)
  (p_1\!\!\!\!\!\slash'+m'_1)\gamma^{\mu}(1-\gamma_{5})
  (p_1\!\!\!\!\!\slash+m_1)  u(\bar{P},S_z),
\end{eqnarray}

where
 \beq
&&m_1=m_c, \qquad m'_1=m_s, \qquad m_2=m_{s}, \qquad m_3=m_{d},\nonumber\\
&&\gamma_{\perp\alpha}=\gamma_{\alpha}-P'\!\!\!\!\!\slash
P'_{\alpha}/M'^2, \qquad
\gamma_{\perp\beta}=\gamma_{\beta}-P\!\!\!\!\slash P_{\beta}/M^2
 \eeq
with $P$($P'$) is the four-momentum of $\Xi_{c}$($\Xi$) and
 $M$($M'$) is the mass of $\Xi_{c}$ ($\Xi$). Setting $\tilde{p}_1=\tilde{p}'_1+\tilde{Q}$,
$\tilde{p}_2=\tilde{p}'_2$ and $\tilde{p}_3=\tilde{p}'_3$ one has
 \beq
 x_{1,2,3}'=x_{1,2,3}, \quad
 k'_{1\perp}=k_{1\perp}-(1-x_1)Q_{\perp}, \quad
 k'_{2\perp}=k_{2\perp}+x_2Q_{\perp}, \quad
 k'_{3\perp}=k_{3\perp}+x_3Q_{\perp}.
 \eeq

The form factors for the weak transition $\Xi_{c}\rightarrow
\Xi$  are defined in the standard way as
\begin{eqnarray}\label{s1}
&& \la \Xi(P',S',S_z') \mid \bar{s}\gamma_{\mu}
 (1-\gamma_{5})c \mid \Xi_{c}(P,S,S_z) \ra  \non \\
 &=& \bar{u}_{\Xi}(P',S'_z) \left[ \gamma_{\mu} f^s_{1}
 -i\sigma_{\mu \nu} \frac{ Q^{\nu}}{M_{\Xi_{c}}}f^s_{2}
 +\frac{Q_{\mu}}{M_{\Xi_{c}}} f^s_{3}
 \right] u_{\Xi_{c}}(P,S_z) \nonumber \\
 &&-\bar u_{\Xi}(P',S'_z)\left[\gamma_{\mu} g^s_{1}
  -i\sigma_{\mu \nu} \frac{ Q^{\nu}}{M_{\Xi}}g^s_{2}+
  \frac{q_{\mu}}{M_{\Xi_{c}}}g^s_{3}
 \right]\gamma_{5} u_{\Xi_{c}}(P,S_z).
\end{eqnarray}
where  $Q \equiv P-P'$. Similarly one can write out the transition matrix of $\Xi'_{c}\rightarrow
\Xi$ whose form factors are denoted as $f_i^v$ and $g_i^v$.

The detailed expressions of the form factors are
\begin{eqnarray}\label{s21}
f^s_1
 &=& -\frac{\sqrt{3}}{2}\int\frac{ d x_2 d^2 k^2_{2\perp}}{2(2\pi)^3}\frac{ d x_3 d^2 k^2_{3\perp}}{2(2\pi)^3}
 \frac{{\rm Tr}[(\bar{P'}\!\!\!\!\!\slash-M_0')\gamma_{5}(p_2\!\!\!\!\!\slash+m_2)(\bar{P}\!\!\!\!\!\slash+M_0)
 \gamma_{5}(p_3\!\!\!\!\!\slash-m_3)]}{\sqrt{M_0^3(m_1+e_1)
 (m_2+e_2)(m_3+e_3)(m_1'+e_1')
 (m_2'+e_2')(m_3'+e_3')}}\nonumber \\
  &&\times\frac{\phi_{\Xi_{c}}^*(x',k'_{\perp})
  \phi_{\Xi_{c}}(x,k_{\perp})}{16\sqrt{x_1x'_1}} \frac{{\rm Tr}[ (\bar{P}\!\!\!\!\!\slash+M_0)\gamma^+(\bar{P'}\!\!\!\!\!\slash+M_0')
  (p_1\!\!\!\!\!\slash'+m'_1)\gamma^{+}
  (p_1\!\!\!\!\!\slash+m_1)  ]}{8P^+P'^+},
\nonumber\\
\frac{f^s_2}{M_{\Xi_{c}}}
 &=& -\frac{\sqrt{3}}{2}\frac{i}{q_{\perp}^i}\int\frac{ d x_2 d^2 k^2_{2\perp}}{2(2\pi)^3}\frac{ d x_3 d^2 k^2_{3\perp}}{2(2\pi)^3}
 \frac{{\rm Tr}[(\bar{P'}\!\!\!\!\!\slash-M_0')\gamma_{5}(p_2\!\!\!\!\!\slash+m_2)(\bar{P}\!\!\!\!\!\slash+M_0)\gamma_{5}(p_3\!\!\!\!\!\slash-m_3)]}{\sqrt{M_0^3(m_1+e_1)
 (m_2+e_2)(m_3+e_3)(m_1'+e_1')
 (m_2'+e_2')(m_3'+e_3')}}\nonumber \\
  &&\times\frac{\phi_{\Xi}^*(x',k'_{\perp})
  \phi_{\Xi_{c}}(x,k_{\perp})}{16\sqrt{x_1x'_1}} \frac{{\rm Tr}[ (\bar{P}\!\!\!\!\!\slash+M_0)\sigma^{i+}(\bar{P'}\!\!\!\!\!\slash+M_0')
  (p_1\!\!\!\!\!\slash'+m'_1)\gamma^+
  (p_1\!\!\!\!\!\slash+m_1)  ]}{8P^+P'^+},
\nonumber\\
g^s_1
 &=& -\frac{\sqrt{3}}{2}\int\frac{ d x_2 d^2 k^2_{2\perp}}{2(2\pi)^3}\frac{ d x_3 d^2 k^2_{3\perp}}{2(2\pi)^3}
 \frac{{\rm Tr}[(\bar{P'}\!\!\!\!\!\slash-M_0')\gamma_{5}(p_2\!\!\!\!\!\slash+m_2)(\bar{P}\!\!\!\!\!\slash+M_0)
 \gamma_{5}(p_3\!\!\!\!\!\slash-m_3)]}{\sqrt{M_0^3(m_1+e_1)
 (m_2+e_2)(m_3+e_3)(m_1'+e_1')
 (m_2'+e_2')(m_3'+e_3')}}\nonumber \\
  &&\times\frac{\phi_{\Xi}^*(x',k'_{\perp})
  \phi_{\Xi_{c}}(x,k_{\perp})}{16\sqrt{x_1x'_1}} \frac{{\rm Tr}[ (\bar{P}\!\!\!\!\!\slash+M_0)\gamma^+\gamma_5(\bar{P'}\!\!\!\!\!\slash+M_0')
  (p_1\!\!\!\!\!\slash'+m'_1)\gamma^{+}\gamma_5
  (p_1\!\!\!\!\!\slash+m_1)  ]}{8P^+P'^+},
\nonumber\\
\frac{g^s_2}{M_{\Xi_{c}}}
 &=&-\frac{\sqrt{3}}{2} \frac{-i}{q_{\perp}^i}\int\frac{ d x_2 d^2 k^2_{2\perp}}{2(2\pi)^3}\frac{ d x_3 d^2 k^2_{3\perp}}{2(2\pi)^3}
 \frac{{\rm Tr}[(\bar{P'}\!\!\!\!\!\slash-M_0')\gamma_{5}(p_2\!\!\!\!\!\slash+m_2)(\bar{P}\!\!\!\!\!\slash+M_0)
 \gamma_{5}(p_3\!\!\!\!\!\slash-m_3)]}{\sqrt{M_0^3(m_1+e_1)
 (m_2+e_2)(m_3+e_3)(m_1'+e_1')
 (m_2'+e_2')(m_3'+e_3')}}\nonumber \\
  &&\times\frac{\phi_{\Xi}^*(x',k'_{\perp})
  \phi_{\Xi_{c}}(x,k_{\perp})}{16\sqrt{x_1x'_1}} \frac{{\rm Tr}[ (\bar{P}\!\!\!\!\!\slash+M_0)\sigma^{i+}\gamma_5(\bar{P'}\!\!\!\!\!\slash+M_0')
  (p_1\!\!\!\!\!\slash'+m'_1)\gamma^+\gamma_5
  (p_1\!\!\!\!\!\slash+m_1)  ]}{8P^+P'^+},
\nonumber\\
f^v_1
 &=&\frac{1}{2}\int\frac{ d x_2 d^2 k^2_{2\perp}}{2(2\pi)^3}\frac{ d x_3 d^2 k^2_{3\perp}}{2(2\pi)^3}
 \frac{{\rm Tr}[\gamma_{\perp}^\alpha(\bar{P'}\!\!\!\!\!\slash'+M_0')\gamma_{5}(p_2\!\!\!\!\!\slash+m_2)(\bar{P}\!\!\!\!\!\slash'+M_0)
 \gamma_{5}\gamma_{\perp}^\beta(p_3\!\!\!\!\!\slash-m_3)]]}{\sqrt{M_0^3(m_1+e_1)
 (m_2+e_2)(m_3+e_3)(m_1'+e_1')
 (m_2'+e_2')(m_3'+e_3')}}\nonumber \\
  &&\times\frac{\phi_{\Xi}^*(x',k'_{\perp})
  \phi_{\Xi_{c}}(x,k_{\perp})}{48\sqrt{x_1x'_1}}  \frac{{\rm Tr}[ (\bar{P}\!\!\!\!\!\slash+M_0)\gamma^+(\bar{P'}\!\!\!\!\!\slash+M_0')
  \gamma_{\perp\alpha}\gamma_{5}(p_1\!\!\!\!\!\slash'+m'_1)\gamma^{+}
  (p_1\!\!\!\!\!\slash+m_1)\gamma_{\perp\beta}\gamma_{5}  ]}{8P^+P'^+},
\nonumber\\
\frac{f^v_2}{M_{\Xi_{c}}}
 &=&\frac{1}{2} \frac{i}{q_{\perp}^i}\int\frac{ d x_2 d^2 k^2_{2\perp}}{2(2\pi)^3}\frac{ d x_3 d^2 k^2_{3\perp}}{2(2\pi)^3}
 \frac{{\rm Tr}[\gamma_{\perp}^\alpha(\bar{P'}\!\!\!\!\!\slash'+M_0')\gamma_{5}(p_2\!\!\!\!\!\slash+m_2)(\bar{P}\!\!\!\!\!\slash'+M_0)\gamma_{5}
 \gamma_{\perp}^\beta(p_3\!\!\!\!\!\slash-m_3)]}{\sqrt{M_0^3(m_1+e_1)
 (m_2+e_2)(m_3+e_3)(m_1'+e_1')
 (m_2'+e_2')(m_3'+e_3')}}\nonumber \\
  &&\times\frac{\phi_{\Xi}^*(x',k'_{\perp})
  \phi_{\Xi_{c}}(x,k_{\perp})}{48\sqrt{x_1x'_1}} \frac{{\rm Tr}[ (\bar{P}\!\!\!\!\!\slash-M_0)\sigma^{i+}(\bar{P'}\!\!\!\!\!\slash-M_0')
\gamma_{\perp\alpha}\gamma_{5}(p_1\!\!\!\!\!\slash'+m'_1)\gamma^{+}
  (p_1\!\!\!\!\!\slash+m_1)\gamma_{\perp\beta}\gamma_{5}   ]}{8P^+P'^+},
\nonumber\\
g^v_1
 &=&\frac{1}{2} \int\frac{ d x_2 d^2 k^2_{2\perp}}{2(2\pi)^3}\frac{ d x_3 d^2 k^2_{3\perp}}{2(2\pi)^3}
 \frac{{\rm Tr}[\gamma_{\perp}^\alpha(\bar{P'}\!\!\!\!\!\slash'+M_0')\gamma_{5}(p_2\!\!\!\!\!\slash+m_2)(\bar{P}\!\!\!\!\!\slash'+M_0)
 \gamma_{5}\gamma_{\perp}^\beta(p_3\!\!\!\!\!\slash-m_3)]}{\sqrt{M_0^3(m_1+e_1)
 (m_2+e_2)(m_3+e_3)(m_1'+e_1')
 (m_2'+e_2')(m_3'+e_3')}}\nonumber \\
  &&\times\frac{\phi_{\Xi}^*(x',k'_{\perp})
  \phi_{\Xi_{c}}(x,k_{\perp})}{48\sqrt{x_1x'_1}} \frac{{\rm Tr}[ (\bar{P}\!\!\!\!\!\slash-M_0)\gamma^+\gamma_5(\bar{P'}\!\!\!\!\!\slash-M_0')
  \gamma_{\perp\alpha}\gamma_{5}(p_1\!\!\!\!\!\slash'+m'_1)\gamma^{+}
  (p_1\!\!\!\!\!\slash+m_1)\gamma_{\perp\beta}\gamma_{5}  ]}{8P^+P'^+},
\nonumber\\
\frac{g^v_2}{M_{\Xi_{c}}}
 &=& \frac{1}{2}\frac{-i}{q_{\perp}^i}\int\frac{ d x_2 d^2 k^2_{2\perp}}{2(2\pi)^3}\frac{ d x_3 d^2 k^2_{3\perp}}{2(2\pi)^3}
 \frac{{\rm Tr}[\gamma_{\perp}^\alpha(\bar{P'}\!\!\!\!\!\slash'+M_0')\gamma_{5}(p_2\!\!\!\!\!\slash+m_2)(\bar{P}\!\!\!\!\!\slash'+M_0)\gamma_{5}
\gamma_{\perp}^\beta(p_3\!\!\!\!\!\slash-m_3)]}{\sqrt{M_0^3(m_1+e_1)
 (m_2+e_2)(m_3+e_3)(m_1'+e_1')
 (m_2'+e_2')(m_3'+e_3')}}\nonumber \\
  &&\times\frac{\phi_{\Xi}^*(x',k'_{\perp})
  \phi_{\Xi_{c}}(x,k_{\perp})}{48\sqrt{x_1x'_1}} \frac{{\rm Tr}[ (\bar{P}\!\!\!\!\!\slash-M_0)\sigma^{i+}\gamma_5(\bar{P'}\!\!\!\!\!\slash-M_0')
  \gamma_{\perp\alpha}\gamma_{5}(p_1\!\!\!\!\!\slash'+m'_1)\gamma^{+}
  (p_1\!\!\!\!\!\slash+m_1)\gamma_{\perp\beta}\gamma_{5}  ]}{8P^+P'^+}.
\end{eqnarray}
These form factors are the same as those in our earlier
paper\cite{Ke:2019smy} except for an additional factor
$-\frac{\sqrt{3}}{2}$ or $\frac{1}{2}$ which exist in corresponding
equations of the equation group (4).

\section{Numerical Results}

\begin{table}
\caption{The quark mass and the parameter $\beta$ (in units of
 GeV).}\label{Tab:t1}
\begin{ruledtabular}
\begin{tabular}{cccccc}
  $m_c$  & $m_s$  &$m_{u}$ & $\beta_{c[su]}$ & $\beta_{s[sq]}$& $\beta_{[cu]}$ \\\hline
  $1.5$  & $0.5$  & 0.25     &0.699     &1.07  & 0.546
\end{tabular}
\end{ruledtabular}
\end{table}

\subsection{The form factors for $\Xi_{c}\to \Xi$ and $\Xi'_{c}\to \Xi$}
In order to study the transitions $\Xi_{c}\to \Xi$ and $\Xi'_{c}\to
\Xi$ one needs to calculate aforementioned form factors numerically
where the parameters in the model need to be pre-determined. The
lifetime of $\Xi_{c}$  and the masses of $\Xi_{c}$, $\Xi'_{c}$ and
$\Xi$ are taken from the data book of particle data
group\cite{Zyla:2020zbs}. The masses of quarks given in
Ref.\cite{Chang:2018zjq} are collected in table I. In fact, we still
know little about the parameters $\beta_1$, $\beta_{23}$, $\beta_1'$
and $\beta_{23}'$ in the wave function for the initial and final
baryons. Generally the reciprocal of $\beta$ is related to the
electrical radius of meson with two constituents. Since the strong
coupling between $q$ and $q^{(')}$ is a half of that between $q\bar
q^{(')}$, for a Coulomb-like interaction if the two interactions are
equal one can expect the electrical separation of $q$ and $q^{(')}$
to be $1/\sqrt{2}$ times that of $q$ and $\bar q^{(')}$ i.e.
$\beta_{qq^{(')}}\approx\sqrt{2}\beta_{q\bar q^{(')}}$. By
considering the binding energy the authors of
Ref.\cite{LeYaouanc:1988fx} obtained the same results. In our early
paper for a compact $q q^{(')}$ system we find
$\beta_{qq^{(')}}=2.9\beta_{q\bar q^{(')}}$. Since the $sq$
subsystem is easy to be broken we think it is not a compact system
and we estimate $\beta_{c[sq]}\approx \sqrt{2}\beta_{c\bar s}$ and
$\beta_{[sq]}\approx \sqrt{2}\beta_{s\bar q}$  where $\beta_{c\bar
s}$ and $\beta_{s\bar q}$ were obtained for the meson
case\cite{Chang:2018zjq}. As for $ss$ we don't know if it is a
compact system so we let  $\beta_{s[sq]}$ be a free parameter to fix
the data of ${\Xi^0}_c\to \Xi^- e^+\nu_e$.
  With these parameters we calculate
the form factors and the transition rates theoretically.

\begin{figure}
\begin{center}
\scalebox{0.6}{\includegraphics{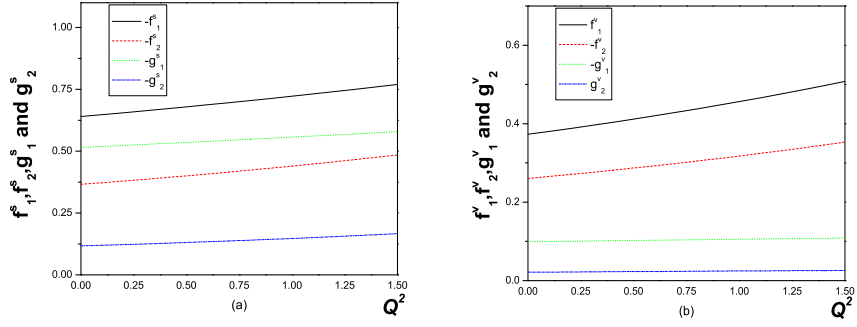}}
\end{center}
\caption{The dependence of  form factors $f^s_1$, $f^s_2$, $g^s_1$
and $g^s_2$ in a three-parameter form on $Q^2$ (a) and the
dependence of the form factors $f^v_1$, $f^v_2$, $g^v_1$ and $g^v_2$
on $Q^2$ (b) .}\label{fg1}
\end{figure}

These  form
factors $f^{s}_i$, $g^{s}_i$, $f^{v}_i$ and $g^{v}_i\,{(i=1,2)}$ are
evaluated in the
space-like region($Q^+=0$ i.e. $Q^2=-Q^2_{\perp}\leq 0$) so should be analytically extrapolated to the time-like
region. In Ref.\cite{pentaquark2} the authors employed a three-parameter form

 \begin{eqnarray}\label{s145}
 F(Q^2)=\frac{F(0)}{
  1-a\left(\frac{Q^2}{M_{\Xi_{c}}^2}\right)
  +b\left(\frac{Q^2}{M_{\Xi_{c}}^2}\right)^2},
 \end{eqnarray}
where $F(Q^2)$ represents the form factors$f^{s}_i$, $g^{s}_i$, $f^{v}_i$ and $g^{v}_i\,{(i=1,2)}$.
 Using the numerical form factors evaluated in the space-like region
we may fix the parameters $a,~b$ and $F(0)$ in the un-physical
region. When one uses Eq. (\ref{s145}) at $Q^2\geq 0$ region these
form factors are extended into the physical region. The values of
$a,~b$ and $F(0)$ for the form factors $f_{1}$, $f_{1}$, $g_{1}$ and
$g_{2}$ are listed in Table \ref{Tab:t1}. The dependence of the form
factors on $Q^2$ is depicted in Fig. \ref{fg1}. The values of $F(0)$
in different works are listed in Table \ref{Tab:t1p} and they
deviate from each other as noted. The differences of $f^s_1$,
$f^s_2$ and $g^s_1$ in
Ref.\cite{Zhao:2021sje,Geng:2020gjh,Zhao:2018zcb,Faustov:2019ddj}
are not very significant but that of $g^s_2$ is. The values of
$f^v_1$, $f^v_2$ , $g^v_1$ and $g^v_2$ at $Q^2=0$ can be found in
Ref.\cite{Azizi:2011mw} and they are very different from ours.

\begin{table}
\caption{The form factors given in the
  three-parameter forms.}\label{Tab:t1}
\begin{ruledtabular}
\begin{tabular}{cccc}
  $F$    &  $F(0)$ &  $a$  &  $b$ \\\hline
  $f^s_1$  &   -0.640    & 0.711    & 0.0981  \\
$f^s_2$  &   -0.366   & 1.07   &0.295   \\
  $g^s_1$  &      -0.515    &  0.471 &  0.0839  \\
  $g^s_2$  &      -0.117  &    1.31  &  0.381  \\
  $f^v_1$  &   0.373    &  1.29    &0.499   \\
$f^v_2$  &   -0.259     &   1.26    & 0.425  \\
  $g^v_1$  &      -0.0990    &    0.458 & 0.312 \\
  $g^v_2$  &      0.0214   &    0.989  &  0.938
\end{tabular}
\end{ruledtabular}
\end{table}

\begin{table}
\caption{The form factors at $Q^2=0$ given in different
 works.}\label{Tab:t1p}
\begin{ruledtabular}
\begin{tabular}{cccccccc}
  $F(0)$    &  this work & Ref.\cite{Zhao:2021sje}  &  Ref.\cite{Geng:2020gjh}&Ref.\cite{Azizi:2011mw}&Ref.\cite{Zhao:2018zcb}&Ref.\cite{Faustov:2019ddj}\\\hline
  $f^s_1(0)$  &   -0.640    & -0.71    & 0.77 &0.194 &-0.567 &0.590\\
$f^s_2(0)$  &   -0.366   &-0.46   &0.96  &0.356 &-0.305&0.441\\
  $g^s_1(0)$  &      -0.515    &  -0.71 &  0.69 &0.311 &-0.491&0.582\\
  $g^s_2(0)$  &      -0.117  &  -0.14 &  0.0068 &0.151&-0.046&-0.184\\
  $f^v_1(0)$  &   0.373     &  -    &-  &0.577&- &-\\
$f^v_2(0)$  &   -0.259     &   -    & -  &0.501&-&-\\
  $g^v_1(0)$  &      -0.0990    &   - & - &0.451&-&- \\
  $g^v_2(0)$  &      0.0214  &    -  & - &0.341&-&-
\end{tabular}
\end{ruledtabular}
\end{table}

\subsection{Semi-leptonic decays of $\Xi_{c} \to
\Xi +l\bar{\nu}_l$ and $\Xi'_{c} \to \Xi +l\bar{\nu}_l$}

\begin{figure}[hhh]
\begin{center}
\scalebox{0.6}{\includegraphics{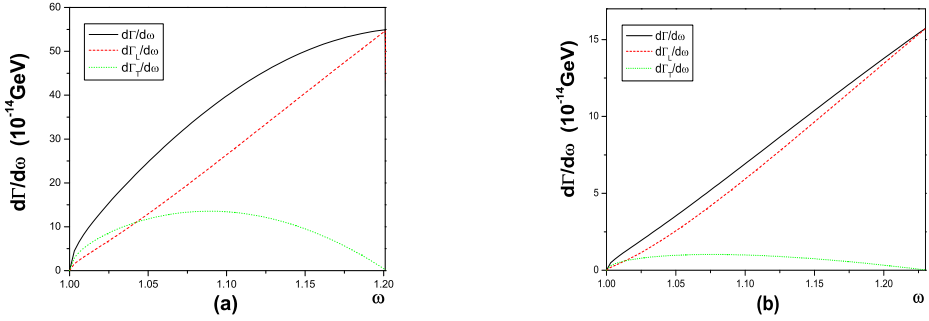}}
\end{center}
\caption{ Differential decay rates $d\Gamma/d\omega$ for the decay
$\Xi_{c} \to \Xi l\bar{\nu}_l$(a) and $\Xi'_{c} \to \Xi
l\bar{\nu}_l$ (b)}\label{f3}
\end{figure}

Using the central value of the data $\Xi^0_{c} \to \Xi^-
e^+\bar{\nu}_e$ from the Belle measurements \cite{Zyla:2020zbs}, we
fix the parameter $\beta_{s[sd]}$. Concretely, we first assign
$\beta_{s[sd]}$ a value and then calculate the form factors $f^s_1$,
$f^s_2$, $g^s_1$ and $g^s_2$ to get the theoretical prediction on
the rate for $\Xi^0_{c} \to \Xi^- e^+\bar{\nu}_e$. Then compare the
theoretical estimates of $BR(\Xi^0_{c} \to \Xi^- e^+\bar{\nu}_e)$
with the data, a deviation would require us to modify the parameter,
repeating the procedure further and further until they are equal to
each other, thus we fix$\beta_{s[sd]}$ to be 1.07 GeV, which is 2.9
times larger than $\beta_{s\bar s}=0.366$ GeV. The value may imply
that the $ss$ pair in $\Xi$ is a more compact subsystem. With the
same parameters the form factors $f^v_1$, $f^v_2$, $g^v_1$ and
$g^v_2$ can be obtained and we evaluate the rate of $\Xi'_{c} \to
\Xi l\bar{\nu}_l$. The differential decay widths $d\Gamma/d\omega$
($\omega=\frac{P\cdot P'}{MM'}$) are shown in Fig. \ref{f3}.

Our numerical results on the decay widths and the ratio of the
longitudinal versus transverse decay rates $R$ (see its expression
in Appendix) of $\Xi^0_{c} \to \Xi^- e^+\bar{\nu}_e$ and $\Xi^+_{c}
\to \Xi^0 e^+\bar{\nu}_e$ are all presented in table \ref{Tab:t4}.
Letting the model parameters (the quark masses and all $\beta$s)
fluctuate up to $\pm$10\%, we estimate the theoretical uncertainties
in our numerical results. Some predictions on the same channels are
also presented in table \ref{Tab:t4}. One may notice the prediction
on $\Gamma(\Xi^0_{c} \to \Xi^- e^+\bar{\nu}_e)$ given by the author
of Ref.\cite{Zhao:2018zcb} is close to our result but his prediction
on the branching ratio is lower than the central value of data
because the measured lifetime of
$\Xi^0_{c}$\cite{PDG18,Zyla:2020zbs} has been significantly modified
since the work \cite{Zhao:2018zcb} was done. The values of $R$ given
in Refs.\cite{Zhao:2018zcb,Zhao:2021sje} is lower than ours. In
table \ref{Tab:t4p} the predictions on $\Xi'_{c} \to \Xi
l\bar{\nu}_l$ are presented. Since the form factors $f^v_1$,
$f^v_2$, $g^v_1$ and $g^v_2$ in Ref.\cite{Azizi:2011mw} deviate from
ours very significantly the width of  $\Xi'_{c} \to \Xi
l\bar{\nu}_l$ they obtained is two orders of magnitude larger than
ours. This should also be tested in more precise measurements in the
future.

\begin{table}
\caption{The theoretical results of $\Xi^0_{c}\to \Xi^-
l\bar{\nu}_l$ (left) and $\Xi^+_{c}\to \Xi^0 l\bar{\nu}_l$
(right).}\label{Tab:t4}
\begin{ruledtabular}
\begin{tabular}{cccc|cccc}
  &  $\Gamma$($10^{-12}{\rm GeV}$) & $\mathcal{B}$&  R  &  $\Gamma$( $10^{-12}{\rm GeV}$)&$\mathcal{B}$ & $R$    \\\hline
 this work&  0.0740$\pm$0.015 & (1.72$\pm$0.35)\%& 2.85$\pm$0.14& 0.0750$\pm$0.016 & (5.20$\pm$1.02)\%&
 2.85$\pm$0.15
\\\hline Ref.\cite{Zhao:2021sje}& $0.145\pm0.031$&$(3.4\pm0.7)\%$ &$1.90\pm0.39$ &$0.147\pm0.032$ &$(10.2\pm2.2)\%$ &$1.90\pm0.38$
\\\hline Ref.\cite{Geng:2020gjh}& -&$(3.49\pm0.95)\%$ &-&- &$(11.3\pm3.35)\%$ &-
\\\hline
Ref.\cite{Azizi:2011mw}&0.4264$\pm$1.490 &$(7.26\pm 2.54)\%$
&-&0.4264$\pm$1.490  &- &- \\\hline Ref.\cite{Zhao:2018zcb}
&0.0791 &$1.35\%$ &1.98 &$0.0803$&$5.39\%$ &1.98
\\\hline Ref.\cite{Faustov:2019ddj}&- &$2.38\%$ &-&- &$9.40\%$ &-

\end{tabular}
\end{ruledtabular}
\end{table}

\begin{table}
\caption{The theoretical results
of
 $\Xi'_{c}\to \Xi
l\bar{\nu}_l$ .}\label{Tab:t4p}
\begin{ruledtabular}
\begin{tabular}{ccc}
   &  $\Gamma$(in unit $10^{-12}{\rm GeV}$) & $R$    \\\hline
 this work&   0.0187$\pm$0.0035  & 10.39$\pm$1.00
 \\\hline
Ref.\cite{Azizi:2011mw}&$1.109$ &-
\end{tabular}
\end{ruledtabular}
\end{table}

\subsection{Non-leptonic decays of ${\Xi_{c}}\to\Xi+ M$ and ${\Xi'_{c}}\to\Xi+ M$}

As a matter of fact, a theoretical exploration for the non-leptonic decays is more complicated than for
semi-leptonic processes.
Based on the
factorization assumption which may be the lowest order approximation by omitting possible final-state-interaction effects, the hadronic transition matrix
element is factorized into a product of two independent hadronic matrix
elements,
\begin{eqnarray}\label{s0}
&& \la \Xi(P',S_z')M \mid \mathcal{H} \mid \Xi^{(')}_{c}(P,S_z) \ra  \nonumber \\
 &=&\frac{G_FV_{cs}V^*_{qq'}}{\sqrt{2}}\la M \mid
\bar{q'} \gamma^{\mu} (1-\gamma_{5}) q \mid 0\ra\la \Xi(P',S_z')
\mid \bar{s} \gamma^{\mu} (1-\gamma_{5})c \mid
\Xi^{(')}_{c}(P,S_z) \ra,
\end{eqnarray}
where the first hadronic matrix element $\la M \mid \bar{q'} \gamma^{\mu} (1-\gamma_{5}) q
\mid 0\ra$ is determined by a well-fixed decay constant and the second one
$\la \Xi(P',S_z')
\mid \bar{s} \gamma^{\mu} (1-\gamma_{5})c \mid
\Xi^{(')}_{c}(P,S_z) \ra$ is evaluated in the previous
sections.
For the decay $\Xi^{(')}_{c}\to\Xi+M$ which is a color-favored channel, the factorization should be a plausible
approximation. The results on these non-leptonic decays can be
checked in the future measurements and the validity degree of the
obtained form factors would be confirmed (within an error range) or suggests serious modifications.

With the theoretical width $\Gamma(\Xi^0_{c}\to\Xi^- \pi^+)$ and the
lifetime of $\Xi^0_c$ one can obtain that branching ratio as
$(1.87\pm0.28)\%$ which is consistent with the measured value of the
Belle collaboration\cite{Li:2018qak} and the up-down asymmetry is
consistent with the present data\cite{Zyla:2020zbs}. From the
results shown in Tab.\ref{Tab:t6}, we find that
$\Gamma(\Xi^0_{c}\to\Xi^- \pi^+)$ and $\Gamma(\Xi^0_{c}\to\Xi^-
\rho^+)$ are very close to each other, but there exists an obvious
gap between their up-down asymmetries. There is a
 similar situation for the decays of $\Gamma(\Xi^0_{c}\to\Xi^- K^+)$
 and $\Gamma(\Xi^0_{c}\to\Xi^- K^{*+})$. One also notices that $\Gamma(\Xi'^0_{c}\to\Xi^-
 M^+)$ is three or four times smaller than $\Gamma(\Xi^0_{c}\to\Xi^-
 M^+)$. The up-down asymmetry for a different channel $\Gamma(\Xi'^0_{c}\to\Xi^-
 M^+)$ is close to 0.5. For $\Gamma(\Xi^+_{c}\to\Xi^0 M^+)$ and $\Gamma(\Xi'^+_{c}\to\Xi^0 M^+)$ their decay
 rates and up-down asymmetry are very close to those of $\Gamma(\Xi^0_{c}\to\Xi^- M^+)$ and $\Gamma(\Xi'^0_{c}\to\Xi^0 M^+)$ which we omit here.
 We estimate the branching ratio of $\Gamma(\Xi^+_{c}\to\Xi^0 \pi^+)$ to be $(5.56\pm0.80)\%$.

\begin{table}
\caption{Our predictions on widths (in unit $10^{-14}{\rm GeV}$)
and up-down asymmetry of non-leptonic decays $\Xi^{(')}_{c}\to\Xi
M$.}\label{Tab:t6}
\begin{ruledtabular}
\begin{tabular}{ccc|ccc}
 mode& width & up-down asymmetry &mode &width  &up-down asymmetry\\\hline
  $\Xi^0_{c}\to\Xi^- \pi^+$ & 8.03$\pm$1.15&-0.975$\pm$0.006&$\Xi'^0_{c}\to\Xi^- \pi^+$  &2.24$\pm$0.36 &0.493$\pm$0.014 \\\hline
  $\Xi^0_{c}\to\Xi^- \rho^+$ &8.53$\pm$1.25&-0.397$\pm$0.013&
  $\Xi'^0_{c}\to\Xi^- \rho^+$&1.93$\pm$0.35&0.557$\pm$0.158\\\hline
  $\Xi^0_{c}\to\Xi^- K^+$   &0.558$\pm$0.086&-0.951$\pm$0.009&$\Xi'^0_{c}\to\Xi^-
  K^+$&0.174$\pm$0.030&0.476$\pm$0.014
                            \\\hline
  $\Xi^0_{c}\to\Xi^- K^{*+}$ &
 0.349$\pm$0.079&-0.252$\pm$0.014&
  $\Xi'^0_{c}\to\Xi^- K^{*+}$&0.0774$\pm$0.0141&0.582$\pm$0.017
\end{tabular}
\end{ruledtabular}
\end{table}

\section{Conclusions and discussions}

In this paper we calculate the transition rate of
$\Xi^{(')}_{c}\to\Xi$ in the light front quark model. For the
baryons $\Xi^{(')}_{c}$ and $\Xi$ we employ the three-quark picture
instead of the quark-diquark one to carry out the calculation. For
$\Xi_c$, the $sq$ content constitutes a physical subsystem which has
a definite spin, instead, in $\Xi$ $ss$ constitutes a physical
subsystem which has also a definite spin. In the transition
$\Xi^{(')}_{c}\to\Xi$, the physical subsystem in the initial state
is different from that in the final state, so that the diquark
picture which is considered as an unchanged spectator cannot be
directly applied in this case. However in the process the strange
quark which does not undergo a change and meanwhile the $q$ ($u$ or
$d$) quark is also approximatively to be a spectator as long as
higher order non-perturbative QCD effects are neglected, so the $sq$
pair is regarded as an effective subsystem in the final state.
Baryon is a three-body system whose total spin can be obtained via
different combinations among that of all three constituents. By a
Racah transformation we can convert one configuration into another.
The Racah coefficients determines the correlation between the two
configuration $(ss)$-$q$ and $s$-$(sq)$. However, one is noted that
the subsystem of $(sq)$ is not regarded as a diquark and moreover
there exists a relative momentum between the two constituents. Thus
in the vertex function of the three-quark picture there exists an
inner degree of freedom for every quark which just manifests by the
relative momentum. In the three-body vertex function of baryon two
quarks are bound to a subsystem which has a definite spin and then
the subsystem couples with the rest quark to form a baryon with the
required spin.

Using the central value of the data $\Xi^0_{c} \to \Xi^-
e^+\bar{\nu}_e$ from the Belle, we fix the parameter $\beta_{s[sd]}$
and calculate the form factors $f^s_1$, $f^s_2$, $g^s_1$, $g^s_2$.
The differences of $f^s_1$, $f^s_2$ and $g^s_1$ given in all the
literatures
Ref.\cite{Zhao:2021sje,Geng:2020gjh,Zhao:2018zcb,Faustov:2019ddj}
are not very significant but that for $g^s_2$ are. We calculate the
ratios of the longitudinal to transverse decay rates $R$ for the
decay $\Xi_{c} \to \Xi e \bar{\nu}_e$ where the definition of $R$ is
given in the appendix, It is noted that the width obtained by Zhao
et al. \cite{Zhao:2018zcb} is close to present data but their
estimated value of $R$ is smaller than ours.

With the same parameters the form factors $f^v_1$, $f^v_2$,
$g^v_1$ and $g^v_2$ can be obtained and they are very different
from those in Ref.\cite{Azizi:2011mw}. We evaluate the decay rate
of $\Xi'_{c} \to \Xi l\bar{\nu}_l$ and the ratios of the
longitudinal to transverse decay rates $R$. The width we obtained
is two order of magnitude smaller than that in
Ref.\cite{Azizi:2011mw}.

Under the factorization assumption, we also evaluate the rates of
several non-leptonic decays. Our numerical results indicate that
$\Gamma(\Xi^0_{c} \to \Xi^- \pi^+) $ and $\Gamma(\Xi^0_{c} \to \Xi^- \rho^+) $
are close to each other but the up-down asymmetries are very
different. The situation for $\Gamma(\Xi^0_{c} \to
\Xi^- K^+) $ and $\Gamma(\Xi^0_{c} \to \Xi^- K^{*+}) $ is very similar.
$\Gamma(\Xi'^0_{c}\to\Xi^-
 M^+)$ is three or four times smaller than $\Gamma(\Xi^0_{c}\to\Xi^-
 M^+)$ but the up-down asymmetry for a different channel $\Gamma(\Xi'^0_{c}\to\Xi^-
 M^+)$ is close to 0.5.

Since there exists a large uncertainty in data, so that none of the
different  theoretical approaches for exploring the semi-leptonic
decay $\mathcal{B}({\Xi^0}_c\to \Xi^- e^+\nu_e)$ can be ruled out so
far. We suggest the experimentalists to make more accurate
measurements on this channel and other non-leptonic decay modes,
thus the data would tell us which approach can work better.
Definitely, the theoretical studies on the baryons are helpful to
understand  the quark model and the non-perturbative QCD effects.

\section*{Acknowledgement}

This work is supported by the National Natural Science Foundation
of China (NNSFC) under the contract No. 12075167, 11975165, 11675082, 11735010, 12035009 and 12075125.

\appendix

\section{Semi-leptonic decays of  $\mathcal{B}_1\to
\mathcal{B}_2  l\bar\nu_l$ }

The helicity amplitudes are expressed with the form factors for
$\mathcal{B}_1\to \mathcal{B}_2 l\bar\nu_l$ through the following
expressions \cite{Korner:1991ph,Bialas:1992ny,Korner:1994nh}
 \beq
 H^V_{\frac{1}{2},0}&=&\frac{\sqrt{T_-}}{\sqrt{Q^2}}\left(
  \left(\Mb+\Mc\right)f_1-\frac{Q^2}{\Mb}f_2\right),\non\\
 H^V_{\frac{1}{2},1}&=&\sqrt{2T_-}\left(-f_1+
  \frac{\Mb+\Mc}{\Mb}f_2\right),\non\\
 H^A_{\frac{1}{2},0}&=&\frac{\sqrt{T_+}}{\sqrt{Q^2}}\left(
  \left(\Mb-\Mc\right)g_1+\frac{Q^2}{\Mb}g_2\right),\non\\
 H^A_{\frac{1}{2},1}&=&\sqrt{2T_+}\left(-g_1-
  \frac{\Mb-\Mc}{\Mb}g_2\right).
 \eeq
where $T_{\pm}=2(P\cdot P'\pm \Mb\Mc)$ and $\Mb\, (\Mc)$
represents $M_{\Xi_{c}}$ ($M_{\Xi}$). The amplitudes for the
negative helicities are obtained in terms of the relation
 \beq
 H^{V,A}_{-\lambda'-\lambda_W}=\pm H^{V,A}_{\lambda',\lambda_W},
  \eeq
where the upper (lower) index corresponds to $V(A)$.
 The helicity
amplitudes are
 \beq
 H_{\lambda',\lambda_W}=H^V_{\lambda',\lambda_W}-
  H^A_{\lambda',\lambda_W}.
 \eeq
 $\lambda_W$ is the helicities of the $W$-boson which can be either $0$ or $1$, which
correspond to the longitudinal and transverse polarizations,
respectively.  The longitudinally ($L$) and transversely ($T$)
polarized rates are
respectively\cite{Korner:1991ph,Bialas:1992ny,Korner:1994nh}
 \beq
 \frac{d\Gamma_L}{d\omega}&=&\frac{G_F^2|V_{cb}|^2}{(2\pi)^3}~
  \frac{Q^2~p_c~\Mc}{12\Mb}\left[
  |H_{\frac{1}{2},0}|^2+|H_{-\frac{1}{2},0}|^2\right],\non\\
 \frac{d\Gamma_T}{d\omega}&=&\frac{G_F^2|V_{cb}|^2}{(2\pi)^3}~
  \frac{Q^2~p_c~\Mc}{12\Mb}\left[
  |H_{\frac{1}{2},1}|^2+|H_{-\frac{1}{2},-1}|^2\right].
 \eeq
where $p_c$ is the momentum of $\mathcal{B}_2$ in the reset frame
of $\mathcal{B}_1$.

 The ratio of the longitudinal to
transverse decay rates $R$ is defined by
 \beq
 R=\frac{\Gamma_L}{\Gamma_T}=\frac{\int_1^{\omega_{\rm
     max}}d\omega~Q^2~p_c\left[ |H_{\frac{1}{2},0}|^2+|H_{-\frac{1}{2},0}|^2
     \right]}{\int_1^{\omega_{\rm max}}d\omega~Q^2~p_c
     \left[ |H_{\frac{1}{2},1}|^2+|H_{-\frac{1}{2},-1}|^2\right]}.
 \eeq

\section{$\mathcal{B}_1\to
\mathcal{B}_2 M$} In general, the transition amplitude of
$\mathcal{B}_1\to \mathcal{B}_2 M$ can be written as
 \beq
 {\cal M}(\mathcal{B}_1\to
\mathcal{B}_2 P)&=&\bar
  u_{\Lambda_c}(A+B\gamma_5)u_{\Lambda_b}, \non \\
 {\cal M}(\mathcal{B}_1\to
\mathcal{B}_2 V)&=&\bar
  u_{\Lambda_c}\epsilon^{*\mu}\left[A_1\gamma_{\mu}\gamma_5+
   A_2(p_c)_{\mu}\gamma_5+B_1\gamma_{\mu}+
   B_2(p_c)_{\mu}\right]u_{\Lambda_b},
 \eeq
where $\epsilon^{*\mu}$ is the polarization vector of the final
vector or axial-vector mesons. Including the effective Wilson
coefficient $a_1=c_1+c_2/N_c$, the form factors in the
factorization approximation are \cite{Korner:1992wi,Cheng:1996cs}
 \beq
 A&=&\lambda f_P(\Mb-\Mc)f_1(M^2), \non \\
 B&=&\lambda f_P(\Mb+\Mc)g_1(M^2), \non\\
 A_1&=&-\lambda f_VM\left[g_1(M^2)+g_2(M^2)\frac{\Mb-\Mc}{\Mb}\right],
 \non\\
 A_2&=&-2\lambda f_VM\frac{g_2(M^2)}{\Mb},\non\\
 B_1&=&\lambda f_VM\left[f_1(M^2)-f_2(M^2)\frac{\Mb+\Mc}{\Mb}\right],
 \non\\
 B_2&=&2\lambda f_VM\frac{f_2(M^2)}{\Mb},
 \eeq
where $\lambda=\frac{G_F}{\sqrt 2}V_{cs}V_{q_1q_2}^*a_1$ and $M$
is the meson mass. Replacing  $P$, $V$ by $S$ and $A$ in the above
expressions, one can easily obtain similar expressions for scalar
and axial-vector mesons .

The decay rates of $\mathcal{B}_1\rightarrow \mathcal{B}_2P(S)$
and up-down asymmetries are\cite{Cheng:1996cs}
 \begin{eqnarray}
 \Gamma&=&\frac{p_c}{8\pi}\left[\frac{(\Mb+\Mc)^2-M^2}{\Mb^2}|A|^2+
  \frac{(\Mb-\Mc)^2-m^2}{\Mb^2}|B|^2\right], \non\\
 \alpha&=&-\frac{2\kappa{\rm Re}(A^*B)}{|A|^2+\kappa^2|B|^2},
 \end{eqnarray}
where $p_c$ is the $\mathcal{B}_2$ momentum in the rest frame of
$\mathcal{B}_1$ and $m$ is the mass of pseudoscalar (scalar). For
$\mathcal{B}_1\rightarrow \mathcal{B}_2 V(A)$ decays, the decay
rate and up-down asymmetries are
 \beq
 \Gamma&=&\frac{p_c (E_{\mathcal{B}_2}+\Mc)}{4\pi\Mb}\left[
  2\left(|S|^2+|P_2|^2\right)+\frac{{e_v}^2}{m^2}\left(
  |S+D|^2+|P_1|^2\right)\right],\non\\
 \alpha&=&\frac{4m^2{\rm Re}(S^*P_2)+2{e_v}^2{\rm Re}(S+D)^*P_1}
  {2m^2\left(|S|^2+|P_2|^2 \right)+\varepsilon^2\left(|S+D|^2+|P_1|^2
  \right) },
 \eeq
where $e_v$ ($m$) is energy (mass) of the vector (axial vector)
meson, and
 \begin{eqnarray}
  S&=&-A_1, \non\\
  P_1&=&-\frac{p_c}{e_v}\left(\frac{\Mb+\Mc}
  {E_{\mathcal{B}_2}+\Mc}B_1+M_{\mathcal{B}_1}B_2\right), \non \\
  P_2&=&\frac{p_c}{E_{\mathcal{B}_2}+\Mc}B_1,\non\\
  D&=&-\frac{p^2_c}{e_v(E_{\mathcal{B}_2}+\Mc)}(A_1- M_{\mathcal{B}_1}A_2).
 \end{eqnarray}

\end{document}